\documentclass[11pt]{article}

\usepackage{amsmath,amsthm,amsfonts,amssymb}
\usepackage{graphicx}
\usepackage[usenames]{color}
\usepackage{hyperref}

\newtheorem{theorem}{Theorem}[section]
\newtheorem{lemma}[theorem]{Lemma}
\theoremstyle{definition}
\newtheorem{algorithm}[theorem]{Algorithm}
\newtheorem{corollary}[theorem]{Corollary}

\newtheorem{definition}[theorem]{Definition}
\newtheorem{example}[theorem]{Example}

\newtheorem{proposition}[theorem]{Proposition}

\def\A{{\mathcal{A}}} 

\def\Pr{\mathcal P} 
\def\s{\sigma} 
\def\eps{\varepsilon}
\def\WC{WC}
\def\N{\mathbb N} 
\def\TM{M} 

\newcommand{\cclass}[1]{\mathbf{#1}}

\newcommand{\abs}[1]{\vert #1 \vert}
\newcommand{\inv}{^{-1}}

\title{Report on Generic Case Complexity}

\author{Robert Gilman\\Alexei G. Miasnikov\\Alexey D. Myasnikov\\Alexander Ushakov}

\date{March 20, 2007}

\begin{document}
\maketitle

\begin{abstract}
This article is a short introduction to generic case complexity, which is a recently developed way of measuring the difficulty of a computational problem while ignoring atypical behavior on a small set of inputs. Generic case complexity applies to both recursively solvable and recursively unsolvable problems.   
\end{abstract}

\tableofcontents

\section{Introduction}

Generic case complexity was introduced a few years ago~\cite{KMSS1} as a way of estimating the difficulty of recursively unsolvable problems in combinatorial group theory. More recently generic complexity and related ideas have proved useful in cryptanalysis of public key systems\cite{MSU}.

Combinatorial group theory has its own computational tradition extending back for more than a century. Almost all computational problems in combinatorial group theory are recursively unsolvable. During the 1990's people working under the leadership of Gilbert Baumslag on the Magnus Computational Group Theory Package~\cite{magnus} noticed that for some difficult problems, simple strategies worked well in practice. Computer scientists had come to similar conclusions earlier in regard to NP-complete problems. The group theoretic version of this theme is developed in~\cite{KMSS1}. The authors define a generic set of inputs and show that for a large class of finitely generated groups the word, conjugacy and membership problems can be solved in linear time on a generic set even though these problems might be recursively unsolvable for the group in question.

The main point here is that it can be convenient and practical to work with a complexity measure which focuses on generic sets of inputs to algorithms and ignores sparse sets of atypical inputs. Generic complexity is close in spirit to the analyses by Smale~\cite{Smale} and Vershik and Sporyshev~\cite{Ver} of the simplex algorithm, and to errorless heuristic case complexity~\cite{BT}. It is also close to average case complexity. Generic complexity is simpler and broader in scope than average case complexity. In addition it is a more direct measure of the performance of an algorithm on most inputs. These points are discussed in Section~\ref{subsec:average-vs-generic}.

This article is an informal introduction to generic case complexity. It is not a complete survey of current developments. We restrict ourselves to polynomial time complexity, and we omit probabilistic algorithms. We also do not discuss generic case completeness or the emerging theory of problems undecidable on generic sets. These topics will be included in a more comprehensive treatment to appear later. 

\section{Computational problems}

\subsection{Decision problems and search problems}
The usual approach to the study of computational problems is to begin with Turing machines and membership problems for formal languages. Membership problems are particular cases of decision problems, that is, computational problems whose answer is ``yes'' or ``no''. Every decision problem can be turned into a membership problem, but this transformation may add complications and even change the nature of the problem. For example representing graphs by words over an alphabet is possible but inconvenient and sometimes misleading~\cite{BGS}. Thus we are led to a more general view.

\begin{definition}\label{de:decision problem}
A decision problem is a pair $\Pr = (L,I)$, where $I$ is a countable set of inputs for $\Pr$ and $L\subset I$ is the positive part of $\Pr$. That is, the answer for input $w\in I$ is ``yes'' if $w\in L$ and ``no'' otherwise. 
\end{definition}

\begin{definition}\label{de:search problem}
A search problem is a pair $\Pr = (R,I\times J)$, where $I$ and $J$ are countable sets, and $R\subset I\times J$ is a binary predicate. Given an element $w\in I$, one is required to find $v\in J$ such that $(w,v)\in R$; that is, such that $R(w,v)$ is true. 
\end{definition}

For example take $J$ to be the integers, $I$ be the set of polynomials with coefficients in $J$, and $R$ the set of pairs $(w,v)$ such that $v$ is a root of $w$. Given a polynomial, $w$, one is required to find an integer  root, $v$, of $w$.

In the preceding example the desired $v$ may not exist. In other search problems one may know in advance that $v$'s always exist and that the only task is to find one. 

When we speak of a problem $\Pr$, we mean either a decision problem or a search problem. 

\subsection{Size functions and stratifications}

Decision and search problems are solved by algorithms and partial algorithms. In general to study the complexity of an algorithm $\A$, one compares the resources spent by $\A$ on input $w$ to the size of $w$. In our case the resource is time, or more precisely, the number of steps required for $\A$ to deal with $w$. We are being informal here. The usual formal definition is stated in terms of the number of steps required by a deterministic Turing machine which implements $\A$.

\begin{definition}\label{de:time}
The time consumed by an algorithm $\A$ on an input $w$ is $T_{\A}(w)$, the number of steps performed by an algorithm $\A$ on the input $w$. If $\A$ is a partial algorithm, then  $T_{\A}$ is a partial function. Its domain is set on inputs on which $\A$ halts.
\end{definition}

\begin{definition}\label{de:size}
A size function for a set $I$ is a map $\sigma:I\to \mathbb N$, the nonnegative integers, such that the preimage of each integer is finite.
\end{definition}

\begin{definition}\label{de:stratification}
A stratification for a set $I$ is an ascending sequence of finite subsets whose union is $I$.
\end{definition}

For each size function $\sigma$ determines a stratification with subsets $\sigma\inv(\{0\})$, $\sigma\inv(\{0, 1\}), \ldots$, and every stratification can be obtained from a size function. The finiteness condition in Definitions~\ref{de:size} and~\ref{de:stratification} will be relaxed later when we discuss computational problems with a probability distribution on the set of inputs.

The choice of the size function depends of course on the problem at hand.
If the input $w$ is a natural number, its size may be taken to be the number of symbols in its representation of to a certain base. For any two integer bases greater than $1$ the corresponding sizes are about the same; they differ by at most a fixed multiplicative factor. However if $w$ is written down in unary notation, that is, as a sequence of $w$ $1$'s, its size will be exponentially greater than its size with respect to the bases greater than $1$. 

Consider another example. An input for the satisfiability problem, SAT, is a boolean expression in conjunctive normal form. There is a standard way to write such an expression as a word over a finite alphabet. A natural choice for input size is the length of that word. But if we are taking the trouble to find difficult instances of SAT, it might be reasonable to take the time needed to generate the word as its size instead. It is also worth noting that converting an arbitrary formula to CNF may increase its length exponentially.

It is easy to obtain surprising complexity bounds by choosing artificial size functions. We do not pursue further here the question of defining precisely what is a reasonable size function; nevertheless we trust the reader will agree that the size functions which appear below are reasonable.

\subsection{Worst case complexity}

\begin{definition}
Let $\A$ be an algorithm, $T_{\A}$ its time function, $I$ the set of inputs, and $\sigma$ a size function for $I$. The worst case complexity of $\A$ with respect to $\sigma$ is the function $WC_\A:N\to N$ defined by $WC_\A(n)=\max_{\sigma(w)\le n} T_{\A}(w)$. 
\end{definition}

We are usually not interested in the precise worst case complexity but rather in estimating its rate of growth. We say that a problem $\Pr$ has polynomial worst case complexity if it is solved by an algorithm $\A$ for which $\WC_\A(n)$ is $O(n^k)$ for some $k$. To define exponential worst case complexity, we replace the condition $O(n^k)$ for some $k$ by $O(2^{n^\varepsilon})$ for some $\varepsilon>0$. We write $\Pr\in\cclass{P}$ and $\Pr\in\cclass{E}$ respectively.

Worst case complexity was the first and is still the most commonly used complexity measure. When an algorithm has low worst case complexity, say $C^w_\A(n)$ is $O(n^2)$, we can be pretty sure that it is practical. But the converse is not true because the worst cases, which determine $WC_\A$, may be rare. This phenomenon has been well known since the 1970's. 

The simplex algorithm for linear programming is frequently used as an example of an algorithm for which hard inputs are rare. The algorithm is used hundreds of times daily and almost always works quickly. But it has been shown by  V.~Klee and G.~Minty \cite{K-M} that there are hard inputs. More precisely, the simplex algorithm is in $\cclass{E} - \cclass{P}$. Khachiyan devised an ingenious polynomial time algorithm for linear programming problems~\cite{Kh}, but the simplex algorithm continues to be widely used because the hard inputs never occur in practice. Vershik  and Sporyshev \cite{Ver} and Smale \cite{Smale} showed independently that the simplex algorithm  runs in linear time on a set of inputs of measure one.

Sometimes we want a problem to be difficult. This is the case when we are looking for a computational problem on which to base a public key cryptosystem. Solving an instance of the problem is equivalent to breaking the cryptosystem for a particular key choice, so we want the computational problem to be hard almost all the time (of course it should be easy if you are in possession of certain secret information, otherwise it would be impossible to decode messages). In this situation the worst case behavior of algorithms is irrelevant.

Worst case complexity is not defined for partial algorithms, because they do not always halt. Nevertheless it can be the case that the nonhalting instances are rare for a partial algorithm just as hard instances can be rare for an algorithm. Coset enumeration (probably the first mathematical procedure to be programmed on a computer) is an example of a partial algorithm which is useful in practice for solving instances of a recursively unsolvable problem, namely whether or not a given finite presentation present a finite group.

In the next section we propose a new complexity measure, generic case complexity, which applies to partial algorithms as well as to algorithms. Subsequently we will discuss the extent to which generic case complexity overcomes the deficiencies of worst case complexity and the relation between generic case and average case complexity.

\section{Generic case complexity}\label{se:generic}

Generic case complexity is an attempt to deal with the fact that worst case complexity can be unsatisfactory when the difficult inputs to an algorithm are sparse and not observable in practice. The main idea is to ignore small sets of difficult inputs and consider the worst case complexity on the remaining large set of more tractable inputs. By large we mean generic as defined below. The treatment here will be generalized when we discuss distributional problems in Section~\ref{se:dist-problems} 

\subsection{Asymptotic density} \label{se:asymptotic}

\begin{definition}\label{de:asymptotic}
Let $I$ be a set of inputs with size function $\sigma$. Define $B_n$, the ball of radius $n$, by $B_n=\{w \mid w\in I, \sigma(w)\le n\}$. A subset $R\subset I$ is said to have asymptotic density $\alpha$, written $\rho(R)=\alpha$, if $\lim_{n\to\infty}\abs{R\cap B_n}/\abs{B_n}=\alpha$ where $\abs X$ denotes the size of a set $X$. If $R$ has asymptotic density $1$, it is called generic; and if it has asymptotic density $0$, it is negligible.  
\end{definition}

Recall from Definition~\ref{de:size} that $\sigma\inv(n)$ is always finite.
Some authors use $\limsup$ rather than $\lim$ in Definition~\ref{de:asymptotic}.

Asymptotic density can be defined using spheres in place of balls. The sphere of radius $n$ is $I_n=\{w \mid w\in I,\, \s(w)= n\}$, that is, the set of inputs of size $n$. We say spherical density and volume density to distinguish the two definitions, and we write $\rho^\circ$ for spherical density.

\begin{lemma}\label{le:spherical density}
Keep the notation of Definition~\ref{de:asymptotic}. If almost all spheres are nonempty and $\rho^\circ(K)$ exists, then so does $\rho(K)$, and the two are equal.
\end{lemma}
 
\begin{proof} Set $x_n =|K \cap B_n|$ and $y_n = |B_n|$. Then
$y_n < y_{n+1}$ for almost all $n$, and $\lim y_n = \infty$. By Stolz's theorem
$$
\rho(K) = \lim_{n \rightarrow \infty} \frac{x_n}{y_n} = \lim_{n
\rightarrow \infty} \frac{x_{n} - x_{n-1}}{y_{n} - y_{n-1}} =
\lim_{n\rightarrow \infty} \frac{|K \cap S_n|}{|S_n|} = \rho^\circ(K).$$
 \end{proof}

\subsection{Convergence rates}\label{se:convergence-rate}

A generic subset of inputs is asymptotically large. Whether it appears large in practice depends on how fast the limit in Definition~\ref{de:asymptotic} converges. 

\begin{definition} 
Let $R$ be a subset of $I$, and suppose that the asymptotic density $\rho(R)$ exists. The function $\delta_R(n)=|R \cap B_n|/|B_n|$ is called the frequency function for $R$. 
\end{definition}

\begin{definition}\label{de:strongly generic}
Suppose $R \subseteq I$ and $\delta_R$ is the density function of $R$. We
say that $R$ has asymptotic density $\rho(R)$ with superpolynomial convergence if $|\rho(R) - \delta_R(n)|$ is $o(n^{-k})$ for every natural number $k$. For short we call a generic set with superpolynomial convergence strongly generic. Its complement is strongly negligible.
\end{definition}

Of course one can introduce exponential convergence, superexponential convergence, etc. In the original papers \cite{KMSS1, KMSS2} strong genericity was reserved for generic sets with exponential convergence,but seems that superpolynomial convergence is fast enough to obtain the same results.

\subsection{Generic case complexity of algorithms and problems}\label{se:generic complexity}

In this section we come to the main notion of the paper.

\begin{definition}\label{de:generic-decidability}
Let $\Pr$ be a problem. A partial algorithm, $\A$, for $\Pr$ generically solves $\Pr$ if the halting set, $H_\A$, of $\A$ is a generic subset of the set of inputs, $I$. In this case we say that $\Pr$ is generically solvable.
\end{definition}

In particular any algorithm for $\Pr$ generically solves $\Pr$. We will see that a generically solvable problem may be recursively unsolvable.

\begin{definition}\label{de:generic-strat-decid}
Let $\Pr$ be a problem with size function $\s$, and let $\A$ a partial algorithm for $\Pr$. A function $f:\N\to\N$ is a generic upper bound for $\A$ if the set $H_{\A,f} = \{w \in I \mid w\in H_\A \mbox{ and } T_\A(w)\le f(\s(w))\}$ is generic in $I$. If $H_{\A,f}$ is strongly generic, then $f$ is a strongly generic upper bound.
\end{definition}

Now we are ready to define generic complexity classes of algorithmic problems.

\begin{definition}\label{de:generic_poly}
A problem $\Pr$ is generically solvable in polynomial time if there exists a partial algorithm for $\Pr$ with a polynomial generic upper bound. If there exists a partial algorithm with a polynomial strongly generic upper bound, then $\Pr$ is strongly generically solvable in polynomial time.
\end{definition}

For short we refer to problems in two classes defined above as generically polynomial and strongly generically polynomial respectively. We denote the two classes by $\cclass{GenP}$ and $\cclass{SGP}$. 

It is clear that one can extend Definition~\ref{de:generic_poly} to other time bounds.

\section{Complexity of unsolvable and NP-complete problems}\label{se:examples}

Analysis of some unsolvable and NP-complete decision problems yields in each case an easy partial algorithm with a generic halting set. We present a few of these results here. A more thorough treatment with complete proofs will appear later.

\subsection{The halting problem} \label{se:halting} 

\begin{theorem}[\cite{HM}]\label{th:halting problem}
The halting problem for deterministic Turing machines with semi-infinite tape and tape alphabet $\{a_0,a_1\}$ is generically solvable in polynomial time; that is, it is in $\cclass{GenP}$. 
\end{theorem}

We do not know whether similar results hold for Turing machines with bi-infinite tapes.

The halting problem is the premier recursively unsolvable problem. For Turing machines with semi-infinite tape extending to the right it is required to decide whether or not a given Turing machine halts when started at the leftmost square of a tape filled with $a_0$'s. The set of inputs, $I$, is the set of Turing machines of the given type. Recall that a Turing machine $\TM$ satisfying the conditions of Theorem~\ref{th:halting problem} may be thought of as a map 
\begin{equation}\label{eq:TM program}
p:\{1,2,\ldots, n\} \times \{a_0,a_1\} \to \{0,1,2,\ldots, n\}\times\{a_0,a_1\} \times \{L,R\}
\end{equation}
where $1$ is the initial state of $\TM$, $0$ is the halting state, and $\{2,\ldots,n\}$ are the other states. The meaning of $p(i,a_r)=(j,a_s,L)$ is that if $\TM$ is in state $i$ and scanning a square containing $a_r$, then its next move is to overwrite $a_r$ with $a_s$, move left on the tape, and go to state $j$. Likewise $p(i,a_r)=(j,a_s,R)$ has the same effect except that $\TM$ moves right instead of left. If $\TM$ reaches state $0$, there are no further moves to make, and $\TM$ halts. As the tape for $\TM$ extends infinitely far to the right but not to the left, $\TM$ may attempt to move off the tape to the left. In this case the move is not completed, and $\TM$ crashes.

The map $p$ may be thought of as a program. The states are the numbers of the instructions, and the instruction, except for instruction number $0$, tells what to do depending on which letter of the tape alphabet is currently being scanned. Instruction $0$ halts the machine.

We take the inputs, $I$, to be the set of programs $p$ defined above, and the size of $p$ is defined to be the the number of non-halting states (of which there must be at least one); $I_n$ is the set of programs with $n$ non-halting states. Here is a polynomial time partial algorithm which decides the halting problem.

\begin{algorithm}\label{al:halting problem}
Input a program $p$\\
Run $p$ until the first time it repeats a state\\
If $p$ halts, say ``Yes''\\
If $p$ crashes, say ''No``\\
Else loop forever.
\end{algorithm}

It would be more informative to say ''Don't know`` than to loop forever, but accommodating this possibility would slightly complicate the definitions in Section~{\ref{se:generic complexity}}. We have opted for simplicity.

Algorithm~\ref{al:halting problem} is obviously polynomial time on its domain and clearly correct. Its domain, $D$, is the set of programs which either halt or crash before repeating a state. It remains only to show that $D$ is generic. We sketch the argument.

Let $D'$ be the the set of programs which crash before repeating a state. Since $D'\subset D$, it is enough to show that $D'$ is generic. 
We can easily count the number of programs in $I_n$, the sphere of radius $n$; $|I_n| = (4n)^{2n}$ for $n\ge 1$. Thus by Lemma~\ref{le:spherical density} we are free to use spherical density, $\rho^{\circ}$ instead of volume density, $\rho$.

Consider the programs in the sphere $I_n$. Half of them have $L$ in their first instruction, and the other half have $R$. Thus half the programs in $I_n$ crash immediately, and the other half move right to square $1$ and transfer from state $1$ to another state. There are $n-1$ non-halting states besides state $1$. Thus the proportion of programs in $I_n$ which do not halt or repeat states at the first step is $\frac{1}{2} + \frac{1}{2}\frac{n-1}{n+1}$. 

Let $C_k\subset I$ be the subset of programs not repeating states or halting within the first $k$ moves. The preceding discussion shows that $\rho^\circ(C_1)=1$. Further analysis yields $\rho^\circ(C_k)=1$ for all $k$. 

Programs in $C_k$ move to a new non-halting state for each of their first $k$ steps. At each of these steps half the remaining programs (those which have not previously crashed) move left on the tape and half move right. 
Thus for $n>k$ the proportion of programs in $C_k\cap I_n$ which do not crash in their first $k$ steps is the same as the fraction of random walks of length $k$ on the integers which start at $0$ and reach only nonnegative numbers. By known results that fraction goes to $0$ as $k$ goes to infinity.
 
Fix $\eps>0$. If $k$ is large enough, the fraction of random walks of length $k$ which avoid negative numbers is less than $\eps$. Thus for $n>k$ the proportion of programs in $C_k\cap I_n$ which do not crash in their first $k$ steps is also less than $\eps$. But for $n$ large enough, $|C_k\cap I_n|/|I_n|> 1-\eps$ because $\rho^\circ(C_k)=1$. Hence the fraction of programs in $I_n$ which crash without repeating a state is at least $(1-\eps)^2$. Consequently $\rho^\circ(D')=1$ as desired.

\subsection{The Post Correspondence Problem}

The set of inputs for the Post Correspondence Problem all finite sequences of pairs of words $(u_1, v_1)\ldots (u_n,v_n)$, $n\ge 1$, over a fixed finite alphabet $\{a_1,\ldots, a_k\}$, $k\ge 2$. The output is ''Yes`` if $u_{i_1}\cdots u_{i_m} = v_{i_1}\cdots v_{i_m}$ for some sequence of indices of length $m\ge 1$ and ''No`` otherwise. We define $I_n$ to be the collection of inputs with $n$ pairs of words of length between $1$ and $n$.

It is well known that PCP is recursively unsolvable. Nevertheless there is a trivial partial algorithm which works well enough to show that PCP is strongly generically polynomial.

\begin{algorithm}\label{al:PCP}
Input an instance of the Post Correspondence Problem\\
If for all $i$, neither $u_i$ nor $v_i$ is a prefix of the other,say ''No``\\
Else loop forever
\end{algorithm}

For any solution $u_{i_1}\cdots u_{i_m} = v_{i_1}\cdots v_{i_m}$ it is clear that one of $u_{i_1}, v_{i_1}$ is a prefix of the other. Thus our algorithm never gives a wrong answer.

\begin{theorem}\label{th:PCP}
The Post Correspondence Problem is strongly generically polynomial; that is, it is in $\cclass{SGP}$.
\end{theorem}

\begin{proof}
The size of $I_n$ is $(1+k+\cdots + k^n)^{2n}$. If we restrict $u_1$ to be a prefix of $v_k$, then there are at most $n+1$ possibilities for $u_1$. Thus the number of inputs in $I_n$ in which $u_1$ is a prefix of $v_1$ is no more than $(n+1)(1+k+\cdots + k^n)^{2n-1}$. We conclude that the number of inputs in $I_n$ for which some $u_i$ is a prefix of $v_i$ or vice-versa is at most $2n(n+1)(1+k+\cdots + k^n)^{2n-1}$. Dividing this number by $|I_n|$ yields $\frac{2n(n+1)}{1+k + \cdots + k^n}$, which approaches $0$ exponentially fast as $n$ goes to infinity.
\end{proof}

\subsection{3-Satisfiability}

SAT has long been known to be easy almost all the time, and there is considerable experimental evidence that 3-SAT is too~\cite{CM}. Thus it is no surprise that 3-SAT is generically easy. 

An instance of 3-SAT, i.e., an input for 3-SAT, is a finite conjunction of clauses
$$
[10'\vee 101 \vee 1] \wedge [110 \vee 11'\vee 111] \wedge \cdots
$$
where the variables are positive integers written in binary, and $'$ denotes
negation. The problem is to decide whether or not there is a truth assignment to the variables which makes all the clauses true.

If the eight different clauses with variables $1, 10, 11$ and their negations all appear in the input, then the formula is not satisfiable. Thus the following partial algorithm is correct.

\begin{algorithm}\label{al:3-sat}
Input an instance of 3-SAT\\
If all the clauses with variables $1, 10, 11$ occur, say ''No``\\
Else loop forever
\end{algorithm}

\begin{theorem}\label{th:3-sat}
 3-SAT is in $\cclass{SGP}$
\end{theorem}

Start with the regular language of clauses 
$$R=[1(0+1)^* (\vee + '\vee) 1(0+1)^* (\vee +'\vee) 1(0+1)^* (]+']).$$ Inputs for 3-SAT are words in the free submonoid $(R\wedge)^*$ of $\Sigma^*$, and size is word length.

Think of inputs as words over the countable alphabet of clauses. To prove Theorem~\ref{th:3-sat} it suffices to show that the set of words which omit some fixed clause is strongly negligible; for then the set of words omitting any of the eight clauses just mentioned will be strongly negligible too. Hence Algorithm~\ref{al:3-sat}, which searches the input for these clauses, will find them in linear time on a generic set of inputs. 

In fact the set of words which omit some fixed clause is asymptotically negligible with exponential convergence. Proof of this fact requires a using the Perron Frobenius Theorem to compare maximum eigenvalue for the incidence matrix of a finite automaton recognizing $R$ with the maximum eigenvalue for the incidence matrix of a finite automaton recognizing the sublanguage of $R$ which omits the eight clauses.

\section{Difficult instances}\label{se:black holes}

Let $\Pr$ be a hard problem, say undecidable or NP-complete. The results of Section~\ref{se:examples} show that difficult instances of $\Pr$ may be rare.  Sometimes we want to find hard instances. For example consider a cryptosystem based on an underlying computational problem $\Pr$. The partial algorithms for $\Pr$ may be viewed as attacks on $\mathcal{C}$; the hard instances are the good keys. 

How do we find hard instances? Typical existing descriptions for NP-complete problems use the notions of parameters and phase-transitions. See for example~\cite[Section~3]{CM} for a discussion of the location of difficult instances of 3-SAT. This approach is good for initial analysis, but quite often further study reveals that the description does not reflect the complexity of the set of hard instances of the problem. Our strong belief is that if the problem $\Pr$ is algorithmically hard then the set of hard instances cannot be satisfactorily described by parameters.

In~\cite{Lynch} Nancy Lynch showed that if $\Pr$ is a decision problem not in $\cclass{P}$, then one can construct a recursive subset of inputs, $J\subset I$, such that for any partial algorithm, $\A$ for $\Pr$ and any polynomial $p$, $\A$ succeeds in time $p(n)$, where $n$ is the size of the input, on only finitely many inputs in $J$. $J$ is called a polynomial complexity core for $\Pr$. 

Lynch's construction involves enumerating all partial algorithms, so it is not practical. In her paper she asks whether certain decision problems might admit a practical construction. As far as we know, none has been proposed.

Lynch's result attracted the interest of many other researchers. See~\cite{YS} for a recent account of subsequent work. The implications of this work for the theory of generic complexity are not yet known.

\section{Distributional computational problems}\label{se:dist-problems}

In this section we generalize the definition of asymptotic density (Definition~\ref{de:asymptotic}) by allowing ensembles $\{\mu_n\}$ of probability distributions. Each $\mu_n$ is a probability distribution on the ball $B_n$ (or sphere $I_n$.) Balls no longer need to be finite, and a subset $R\subset I$ has volume density $\alpha$ if $\lim \mu_n(R\cap B_n) = \alpha$. Spherical density is defined similarly. The discussion of generic case complexity in Section~\ref{se:generic} makes sense with the generalized definition of asymptotic density in place of the original definition. The original definition corresponds to the case of uniform distributions on finite balls and spheres. We write $\rho_\mu$ and $\rho^\circ_\mu$ for volume density and spherical density defined with respect to the measure $\mu$.

One source of ensembles $\{\mu_n\}$ is probability distributions on $I$. Given a probability distribution, $\mu$, on $I$, we define $\mu_n$ for each $n$ to be the conditional probability on $B_n$ or $I_n$. We assume that $\mu$ is atomic, i.e., that $\mu(x)$ is defined for every singleton $\{x\}$. For any subset $R \subset I$, $\mu(R) = \sum_{x \in R} \mu(x)$. 

\begin{definition} 
A distributional computational problem is a pair $(\Pr,\mu)$ where $\Pr$ is a computational problem and $\mu$ is a probability measure on $I$. 
\end{definition}

Here is an example to illustrate how $\mu$ might arise in practice. Consider the following search problem from combinatorial group theory. For a fixed finite presentation of a group $G$, the set of inputs, $I$, consists of all words (in the generators of $G$) defining the identity in $G$. It is required for each $w\in I$ to verify that $w$ does define the identity by constructing a certain kind of proof, and a particular procedure is introduced for that purpose. The details are not important here, the point is that $I$ is recursively enumerable but need not be recursive. How then are we to define a reasonable stratification in order to estimate the generic complexity of our procedure? Stratifying by the length of $w$ is not useful because the resulting $B_n$'s need not be recursive. The answer~\cite{Sasha} is to define a random walk over $I$ which stops at each point in $I$ with positive probability and thus induces $\mu$. 

\subsection{Average case complexity} \label{se:average-case}

Average case complexity provides a measure of the difficulty of a distributional problem. The definition of average case complexity was motivated by the observation that some NP-complete problems admit algorithms which seem to run quickly in practice. The idea was to explain this phenomenon by showing that although in the maximum running time of an algorithm $\A$ over all inputs of size $n$ might be very high, the average running time might be much smaller. For this purpose the NP-complete problem was converted to a distributional problem by introducing a probability distribution $\mu$ on the set of inputs, $I$.

As average case complexity is very similar to generic case complexity, we will discuss the relation between the two in some detail. We begin with a quick review of average case complexity. We refer the reader to Levin's original paper~\cite{Le}, which has been further developed by Gurevich~\cite{Gu} and Impagliazzo~\cite{Impagliazzo05}.

\begin{definition} An algorithm $\A$ is polynomial time on $\mu$-average if its time function, $T_\A$ satisfies $T_\A(x) \leq f(x)$ for some polynomial on $\mu$-average function $f$. The class of distributional problems decidable in time polynomial on average is denoted by $\cclass{AvP}$.
\end{definition}

It remains to define when a function is polynomial on average. A straightforward definition would be the following.

\begin{definition}
\label{de:Gu-spheres}
A function $f:I \rightarrow \mathbb{R}^+$ is expected polynomial on spheres (with respect to an ensemble of spherical distributions $\{\mu_n\}$) if there exists $k \geq 1$ such that
   \begin{equation} \label{eq:naive-poly-ave}
   \int_{I_n} f(w)\mu_n(w) = O(n^k).
    \end{equation}
\end{definition}

However, in order to obtain closure under addition, multiplication, and multiplication by a scalar we must define a larger class. 

\begin{definition}[\cite{Le}]\label{de:Levin} 
A function  $f:I \rightarrow \mathbb{R}^+$ is polynomial on $\mu$-average if there exists $\varepsilon  > 0$ such that $\int_{I} (f(x))^\varepsilon \s(x)^{-1} \mu(x) < \infty$.
\end{definition}

\noindent
Which is equivalent to the following.

\begin{definition}[\cite{Impagliazzo95}]\label{de:Impagliazzio}
Let $\{\mu_n\}$ be an ensemble of volume distributions on balls $\{B_n\}$ of $I$. A function $f:I \rightarrow \mathbb{R}$ is polynomial on average  with respect to $\{\mu_n\}$ if there exists an $\varepsilon > 0$ such that $\int_{B_n} f^\varepsilon(x) \mu_n(x) = O(n)$.
\end{definition}

\subsection{Average Case vs Generic Case}
 \label{subsec:average-vs-generic}

Average case complexity provides a more balanced assessment of the difficulty of an algorithmic problem than worst-case complexity. Many algorithmic problems, such as the NP-complete Hamiltonian Circuit Problem~\cite{GurevichShelah}, are hard in the worst case but easy on average for reasonable distributions.

Average case complexity is very similar to generic case complexity, but we argue that the latter has certain advantages. Generic complexity applies to undecidable problems as well as to decidable problems, it is easier to employ than average complexity, and it is a direct measure of the difficulty of a problem on most inputs. Average case complexity tells us something else. In \cite{Gu2} Gurevich explains, in terms of a Challenger-Solver game, that average case analysis describes the fraction of hard instances of the problem with respect to a measure of difficulty. In other words to have polynomial on average time an algorithm should have only a sub-polynomial fraction of inputs that require superpolynomial time to compute.

Now we give some more precise comparisons. Our first observation is that $\cclass{AvP}$ and  $\cclass{GenP}$ are incomparable; that is, $\cclass{GenP} - \cclass{AvP}$ and $\cclass{AvP} - \cclass{GenP}$ both contain functions. We leave it as an exercise for the reader to verify the first assertion by constructing a function which is very large on a negligible set and small on the complementary generic set. The second part follows from the next example.

\begin{example}
Let $I = \{0, 1\}^*$. For $w\in I$ define $\s(w)=\abs w$, the length of $w$, and define $\mu(w) = 2^{ - 2|w| - 1}$. Consider  $f : I \to N$ defined by $F(w)=2^{\abs w}$. Observe that $f\in \cclass{AvP}$ by Definition~\ref{de:Levin} with $\varepsilon < 1$, but $f\notin \cclass{GenP}$.
\end{example}

However, a big chunk of $\cclass{AvP}$ does lie in $\cclass{GenP}$, namely the functions satisfying Definition~\ref{de:Gu-spheres}. 

\begin{proposition}\label{pr:aver-impl-gen}
If a function $f:I \rightarrow \mathbb{R}^+$ is polynomial on $\mu$-average on spheres, then $f$ is generically polynomial relative to the asymptotic density $\rho_\mu$.
\end{proposition}

\begin{proof}
If $f$ is an expected polynomial then there exists a constant $c$
and $k \geq 1$  such that for any $n$
$$\int_{I_n} f^{\frac{1}{k}}(w) \mu_n(w) \leq c n.$$

It follows that  for any polynomial $q(n)$ $$\mu_n\{ x \in I_n\mid f^{\frac{1}{k}}(x)> q(n) cn\} \le 1/q(n),$$ Now let  $S(f,q,k) = \{x \in I \mid f(x) \geq \left (cq(s(x))s(x) \right )^k\}$ be the set of those instances from $I$ on which $f(x)$ is not bounded by  $\left (cq(s(x))s(x) \right )^k$. Then $$\mu_n(I_n \cap S(f,q,k))= \mu_n\{ x \in I_n\mid f^{\frac{1}{k}}(x)> q(n) cn\} \leq 1/q(n),$$ therefore, the asymptotic density $\rho_\mu$ of $S(f,q,k)$ exists and equal to $0$.  This shows that $f$ is generically bounded by the polynomial $(cq(n)n)^k$. 
\end{proof}

Proposition \ref{pr:aver-impl-gen}  gives a large class of polynomial on average functions which are generically polynomial.

\begin{corollary}
Let $\A$ be an algorithm for the distributional problem $\Pr$.
If the expected time of $\A$ with respect to the spherical
distributions is bounded above by a polynomial then $\A\in \cclass{GenP}$. 
\end{corollary}

On the other hand under some conditions membership in $\cclass{GenP}$ implies membership in $\cclass{AvP}$. We refer the reader to~\cite{KMSS2}.


\begin{thebibliography}{\hspace{0.5in}}

\bibitem{BGS} A. Blass, Y. Gurevich and S. Shelah, On Polynomial Time Computation Over Unordered Structures
Journal of Symbolic Logic 67 (2002), 1093-1125.

\bibitem{BT} A. Bogdanov and L. Trevisan, {\em Average-Case Complexity}, Nowpublishers, 2006.

\bibitem{BD} Book, Ronald V., and Du, Ding Zhu,
The existence and density of generalized complexity cores.
J. Assoc. Comput. Mach. 34 (1987), no. 3, 718--730. 

\bibitem{CM} S. A. Cook and D. G. Mitchell, Finding hard instances of the satisfiability problem: a survey, in {\em Satisfiability Problem: Theory and Applications}, D. Du, J. Gu, and P. M. Pardalos, eds., DIMACS Series in Discrete Mathematics and TheoreticalComputer Science, {\bf 35} 1997, 1--17.

\bibitem{Cooper}  S.B.Cooper, {\it Computability Theory}, Chapman
and Hall/CRC Mathematics, 2003.

\bibitem{GJ} M. Garey and J. Johnson, {\em Computers and
  Intractability, A Guide to NP-Completeness}, W. H. Freeman, 1979

\bibitem{Gu} Y. Gurevich, \emph{Average case completeness},
J. of Computer and System Science  \textbf{42} (1991), 346--398.

\bibitem{Gu2} Y. Gurevich, {\em The Challenger-Solver game:
Variations on the theme of P =?NP},   Logic in Computer Science
Column, The Bulletin of EATCS, October 1989, p.112-121.

\bibitem{GurevichShelah} Y. Gurevich and S. Shelah. {\em Expected computation time for Hamiltonian Path Problem}, SIAM J. on Computing 16:3 (1987) p.
486-502.

\bibitem{HM} J.D. Hamkins, A.
Miasnikov. The halting problem is decidable on a set of asymptotic
probability one. Notre Dame Journal of Formal Logic, 47 (2006), No.
4, 515--524.

\bibitem{Impagliazzo95} R. Impagliazzo, {\em A personal view of
average-case complexity}, Preprint, 1995.

\bibitem{Impagliazzo05} R. Impagliazzo,  {\em Computational Complexity Since 1980},
Springer-Verlag Berlin Heidelberg, R. Ramanujam and S. Sen (Eds.):
FSTTCS 2005, LNCS 3821,  2005, pp. 1947.

\bibitem{Kh} L. Khachiyan, A polynomial algorithm in linear programming. Dokl.\ Akad.\ Nauk SSSR 244 (1979), no. 5, 1093--1096. \{English translation: Soviet Math.\ Dokl.\ 20 (1979), no. 1, 191--194.\}

\bibitem{K-M}
V.~Klee and G.~Minty, How good is the simplex algorithm? in 
{\em Inequalities, III (Proc. Third Sympos., UCLA)}, 1969, pp. 159--175. Academic Press, New York, 1972.

\bibitem{KMSS1}  I.Kapovich, A.Myasnikov, P.Schupp, V.Shpilrain {\it
Generic-case complexity and decision problems in group theory} J. of
Algebra, 264 (2003), 665-694.

\bibitem{KMSS2} I. Kapovich, A. Myasnikov, P. Schupp, V. Shpilrain {\it
Average-case complexity for the word and membership problems in
group theory}. Advances in Mathematics 190 (2005), 343-359.

\bibitem{Le} L. Levin, Average case complete problems, SIAM
Journal of Computing \textbf{15} (1986), 285--286.

\bibitem{Lynch} N. Lynch, On reducibility to complex or sparse sets.  J. Assoc.\ Comput.\ Mach.\  22  (1975), 341--345. 

\bibitem{magnus} {\em http://sourceforge.net/projects/magnus}

\bibitem{MSU} A. Myasnikov, V. Shpilrain, A. Ushakov, A Practical Attack on a Braid Group Based Cryptographic Protocol, in {\em Lecture Notes in Computer Science}, \textbf{3621}, Springer Verlag, 2005, 86--96.

\bibitem{Smale} S. Smale, {\em On the average number of steps of the simplex method of linear programming}, Mathematical Programming,  27 (1983), pp. 241--262.

\bibitem{Sasha} A. Ushakov, Dissertation, City University of New York, 2005.

\bibitem{Ver} A. M. Vershik, P. V. Sporyshev, {\em An estimate of the average number of steps in the simplex method, and problems in asymptotic integral geometry}. Dokl. Akad. Nauk SSSR 271, No.5, 1044-1048 (1983). English translation: Sov. Math. Dokl. 28, 195-199 (1983).

\bibitem{YS} T. Yamakami, T. Suzuki, Resource bounded immunity and simplicity,
Theoret.\ Comput.\ Sci.\ 347 (2005), 90--129. 

\end{thebibliography}
\end{document}